\documentclass[preprint2]{aastex631}
\usepackage{multirow}
\usepackage{tikz}

\begin{document}

\title{JWST transmission spectroscopy of HD\,209458b: a super-solar metallicity, a very low C/O, and no evidence of CH$_4$, HCN, or C$_2$H$_2$}

\author[0000-0002-6215-5425]{Qiao Xue}
\affiliation{Department of Astronomy \& Astrophysics, University of Chicago, Chicago, IL, USA}
\affiliation{School of Physics and Astronomy, Shanghai Jiaotong University, Shanghai, CN}

\author[0000-0003-4733-6532]{Jacob L.\ Bean}
\affiliation{Department of Astronomy \& Astrophysics, University of Chicago, Chicago, IL, USA}

\author[0000-0002-0659-1783]{Michael Zhang}
\affiliation{Department of Astronomy \& Astrophysics, University of Chicago, Chicago, IL, USA}

\author[0000-0003-0156-4564]{Luis Welbanks}
\affiliation{School of Earth \& Space Exoploration, Arizona State University, Tempe, AZ 85287, USA}

\author{Jonathan Lunine}
\affiliation{Department of Astronomy, Cornell University, Ithaca, NY, USA}

\author[0000-0003-3829-8554]{Prune August}
\affiliation{Department of Astronomy \& Astrophysics, University of Chicago, Chicago, IL, USA}

\begin{abstract}
We present the transmission spectrum of the original transiting hot Jupiter HD\,209458b from 2.3 -- 5.1\,$\mu$m as observed with the NIRCam instrument on the James Webb Space Telescope (JWST). Previous studies of HD\,209458b's atmosphere have given conflicting results on the abundance of H$_2$O and the presence of carbon- and nitrogen-bearing species, which have significant ramifications on the inferences of the planet's metallicity (M/H) and carbon-to-oxygen (C/O) ratio. We detect strong features of H$_2$O and CO$_2$ in the JWST transmission spectrum, which when interpreted using a retrieval that assumes thermochemical equilibrium and fractional grey cloud opacity yields $3^{+4}_{-1}$ $\times$ solar metallicity and C/O = $0.11^{+0.12}_{-0.06}$. The derived metallicity is consistent with the atmospheric metallicity-planet mass trend observed in solar gas giants. The low C/O ratio suggests that this planet has undergone significant contamination by evaporating planetesimals while migrating inward. We are also able to place upper limits on the abundances of CH$_4$, C$_2$H$_2$ and HCN of log($\chi_{\mathrm{CH}_4}$) = -5.6, log($\chi_{\mathrm{C}_2\mathrm{H}_2}$) = -5.7, and log($\chi_{\mathrm{HCN}}$) = -5.1, which are in tension with the recent claim of a detection of these species using ground-based cross-correlation spectroscopy. We find that HD\,209458b has a weaker CO$_2$ feature size than WASP-39b when comparing their scale-height-normalized transmission spectra. On the other hand, the size of HD\,209458b's H$_2$O feature is stronger, thus reinforcing the low C/O inference.
\end{abstract}

%The elemental ratios in the atmosphere of hot jupiters may help us understand planet formation and evolution. In this paper, we used forward model to analyze the transmission spectrum of HD\,209458b obtained by \textit{James Webb Space Telescope} (JWST) NIRCam. We found ... H2O, CO2. Water abundance is above solar. Contrary to findings of \cite{giacobbe_five_2021}, we didn't find HCN and CH4.

%% Keywords should appear after the \end{abstract} command. 
%% The AAS Journals now uses Unified Astronomy Thesaurus concepts:
%% https://astrothesaurus.org
%% You will be asked to selected these concepts during the submission process
%% but this old "keyword" functionality is maintained in case authors want
%% to include these concepts in their preprints.

%% From the front matter, we move on to the body of the paper.
%% Sections are demarcated by \section and \subsection, respectively.
%% Observe the use of the LaTeX \label
%% command after the \subsection to give a symbolic KEY to the
%% subsection for cross-referencing in a \ref command.
%% You can use LaTeX's \ref and \label commands to keep track of
%% cross-references to sections, equations, tables, and figures.
%% That way, if you change the order of any elements, LaTeX will
%% automatically renumber them.
%%
%% We recommend that authors also use the  natbib \citep
%% and \citet commands to identify citations.  The citations are
%% tied to the reference list via symbolic KEYs. The KEY corresponds
%% to the KEY in the \bibitem in the reference list below. 

\section{Introduction} \label{sec:intro}
Detected initially by the radial velocity method in 1999, the hot Jupiter HD\,209458b was the first exoplanet found to transit its host star \citep{charbonneau_detection_2000, henry_transiting_2000}. Since then, it has been one of the most frequently studied exoplanets and it has been the subject of a number of breakthroughs that sparked the study of exoplanetary atmospheres. Via transmission spectroscopy, it was the subject of the first exoplanet atmosphere detection \citep{charbonneau_detection_2002}, the first found to have an escaping atmosphere \citep{vidal-madjar_extended_2003}, the first observed to possess atomic carbon and oxygen \citep{vidal-madjar_detection_2004}, and the first to have its orbital velocity measured, thus turning the system into a double-lined spectroscopic binary \citep{snellen_orbital_2010}. It was one of the first two exoplanets with detected infrared emission \citep{deming_infrared_2005}, and it was one of the first two exoplanets with the spectroscopic identification of water \citep{deming_infrared_2013}. What's more, it was once thought to possess a stratospheric temperature inversion \citep{knutson_3680_2008, burrows_theoretical_2007}, although that hypothesis was later refuted by subsequent observations and analysis \citep{diamond-lowe_new_2014,schwarz_evidence_2015,line_no_2016}.

HD\,209458b is still among the best targets for atmospheric study in this era of thousands of known transiting planets. It has a relatively bright host star, a favorable planet-to-star radius ratio, a high equilibrium temperature, and a low surface gravity. Ultimately, it has the highest transmission spectroscopy metric ($\sim 900$\footnote{https://tess.mit.edu/science/tess-acwg/}) of all known exoplanets \citep{kempton18}.

% It has a relatively bright host star (G0V, $T_{eff}=6076K$), big size ($R_p=1.39M_J$) \citep{gaia_collaboration_gaia_2018}, and favorable $R_p/R_*$ thus easier to observe. It orbits very close to its host star, resulting in high temperatures ($T_{eq} = 1448 K$) \cite{barstow_consistent_2016} and low density, thus an extended, bloated atmosphere.

Despite its great observability, several fundamental questions about the composition of HD\,209458b's atmosphere remain unsolved. One of these questions is its atmospheric water abundance. Several groups have derived sub-solar water abundances from measurements of its atmosphere \citep{madhusudhan_h_2014,macdonald_hd_2017, brogi_framework_2017, welbanks_mass-metallicity_2019, pinhas_h_2o_2019}, with a recent re-analysis providing estimates consistent with solar expectations within the uncertainties \citep{Welbanks2021}. This could be caused by an overall low metallicity or just a low oxygen abundance in the planet, neither of which are expected by standard models for giant planet formation \citep{oberg_2016, booth_2017_chemical}. However, \citet{line_no_2016}, \citet{sing_a_continuum_2016}  and \citet{tsiaras_population_2018} also reported solar to super-solar H$_2$O abundances, which is more in line with expectations from traditional planet formation models \citep{owen_2006_compositional,_berg_2011}.

A second open question surrounds the carbon and nitrogen chemistry. Interpretation of the \textit{Hubble Space Telescope} (\textit{HST}) transmission spectrum initially suggested strong evidence for NH$_3$ and/or HCN \citep{macdonald_hd_2017}, but a subsequent analysis lowered the detection significance and highlighted NH$_3$ as being the more likely of the two \citep{macdonald_signatures_2017}. One the other hand, two high-resolution spectroscopy (HRS) analyses both claimed the presence of HCN \citep{hawker_evidence_2018, giacobbe_five_2021}. The latter of these two studies also claimed the detection of NH$_3$, CH$_4$, and C$_2$H$_2$ in addition to H$_2$O and CO \citep{giacobbe_five_2021}. When assuming equilibrium chemistry and a clear atmosphere, the presence of all of these molecules together suggests a highly sub-solar metallicity ($<$1\% of solar) and/or a C/O ratio of around 1 or higher \citep[compare to the solar value of 0.59,][]{asplund21}, which together challenge planet formation models \citep{mousis_nebular_2012, madhusudhan_h_2014}.

%Nitrogen chemistry serves as powerful diagnostics of disequilibrium atmospheric chemistry and planet formation processes. Two nitrogen-bearing molecules, NH$_3$ and HCN have been reported in its atmosphere,  although the discovery is still debatable.

%A second open question surrounds the carbon chemistry in HD\,209458b's atmosphere is the presence of methane (CH$_4$). Using emission spectroscopy, CH$_4$ has been detected \citep{swain_water_2009} and with high-resolution spectroscopy (HRS), its abundance was constrained to $10^{-3}$ \citep{giacobbe_five_2021}. However, several studies obtained abundance of $10^{-8}$ and left this molecule undetected \citep{line_no_2016, macdonald_hd_2017, brogi_framework_2017, pinhas_h_2o_2019}.

In this paper, we present the first transmission spectrum of the archetypical hot Jupiter HD\,209458b obtained with the \textit{James Webb Space Telescope} (\textit{JWST}) to answer these longstanding questions. We describe our observations and data analysis in \S\ref{sec:observation}, atmospheric modeling in \S\ref{sec:modeling} and results in \S\ref{sec:results}.

\section{Observations and Data Analysis} \label{sec:observation}

\begin{figure*}[ht!]
\includegraphics[width = 18.3 cm]{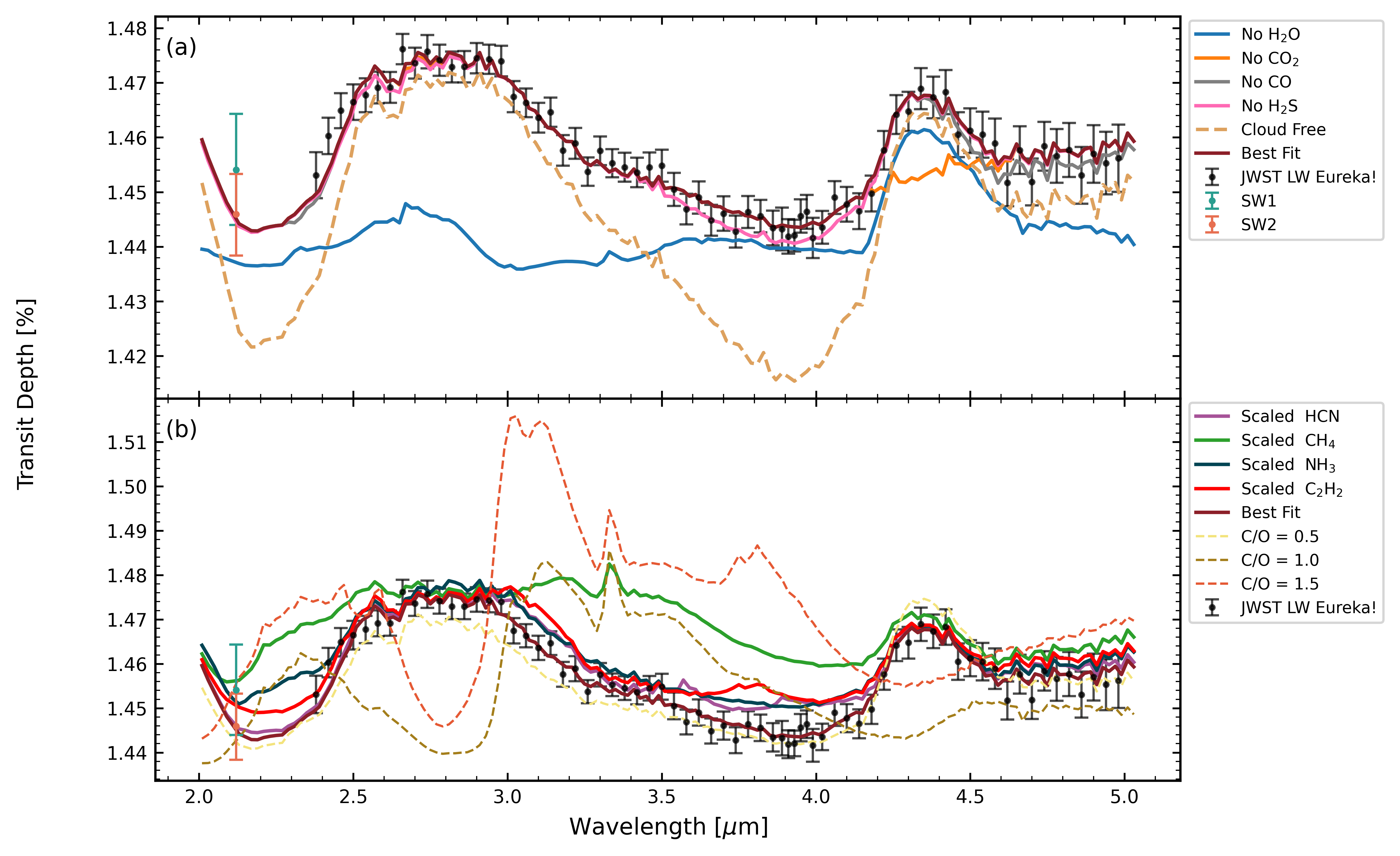} \\
\caption{(a): JWST NIRCam data reduced by \texttt{Eureka!} (points with error bars). The short-wavelength data at 2.12\,$\mu m$ are plotted but not included in the retrieval. The best-fit chemical equilibrium model is shown as the solid maroon line. Absorption contributed by H$_2$O, CO$_2$, CO, H$_2$S and cloud are highlighted. (b): Model spectra of HD\,209458b when scaling the abundances of HCN, CH$_4$, NH$_3$, C$_2$H$_2$ to artificially higher levels are plotted with solid lines (see \S\ref{sec:results} for more details). Models with different C/O are plotted with dashed lines. The data behind the figure can be found in \S\ref{data_availability}.
\label{fig:spectrum_and_fit}}
\end{figure*}

 \begin{table*}[ht!]
  \centering
  \caption{White light curve best-fit parameters by \texttt{Eureka!} and \texttt{SPARTA}. }
  \label{tab:lc_best_fit}
  \centering
  \begin{tabular}{c c c c c}
      \hline
        & \multicolumn{2}{c}{Visit 1} & \multicolumn{2}{c}{Visit 2} \\
      & \texttt{Eureka!}& \texttt{SPARTA} & \texttt{Eureka!} & \texttt{SPARTA} \\
      \hline
      $t_0$ (MJD) & $59889.72649 \pm 0.00003$ 
      & $59889.72646 \pm 0.00003$
      & $59893.25125 \pm 0.00002$ 
      & $59893.25120 \pm 0.00003$ \\
      $a/R_*$ & / & /  & $8.84 \pm 0.02$ & $8.84 \pm 0.02$  \\
      $i$ ($\degr$) & / & / & $86.74 \pm 0.04$ & $86.74 \pm 0.03$\\
      \hline
  \end{tabular}
\end{table*}
We observed two transits of HD\,209458b with JWST's NIRCam instrument \citep{greene__2017} on November 10 and 16, 2022 (program GTO 1274, J.~Lunine PI). Each observation lasted 8.01 hours, which is long enough to sample the 3.12-hour transit and the baseline flux before and after the transit. Both observations used the module A grism R mode to obtain time-series near-infrared spectra in the long-wavelength (LW) channel. The first observation collected data with the F444W filter, yielding spectra from 3.86 to 5.06\,$\mu m$. The second observation used the F322W2 filter, yielding spectra from 2.36 to 4.02\,$\mu m$.

Both visits employed the SUBGRISM64 subarray and BRIGHT2 readout pattern. The first observation used six groups per integration for a total of 7,107 integrations, while the second used four groups per integration for a total of 9,473 integrations. Photometry was also obtained simultaneously in the short-wavelength (SW) channel during each LW channel observation. The SW photometry for both visits was obtained in a narrow band at 2.12\,$\mu m$ using the WLP4 filter with the BRIGHT1 readout pattern. 

We reduced and analyzed the LW data independently using two different pipelines. The first pipeline we used is \texttt{Eureka!} \citep{bell_eureka_2022}, which has been utilized extensively by the \textit{JWST} Transiting Exoplanet Early Release Science (JTEC ERS) Team \citep{w39_co2, ahrer_nircam_2023, alderson_g395h_2023, feinstein_niriss_2023, rustamkulov_prism_2023, coulombe23}. The second pipeline we used is \texttt{SPARTA}\footnote{\url{https://github.com/ideasrule/sparta}}, which was first utilized for the MIRI/LRS phase curve of GJ\,1214b \citep{kempton_reflective_2023} and is also now being included in the JTEC ERS analyses of MIRI/LRS data of WASP-43b (Bell et al., submitted) and WASP-39b (Powell et al., submitted).

Our \texttt{Eureka!} implementation follows the analyses of NIRCam data presented in \citet{ahrer_nircam_2023} and \citet{bean_high_2023}. The first two stages in \texttt{Eureka!} are identical with those in the JWST Science Calibration Pipeline (\texttt{jwst}) except that we did a group-level background subtraction and we increased the jump detection threshold to 6.0. Following Stage 2, \texttt{Eureka!} performed a column-by-column linear fit to calculate the background in the region beyond 13 pixels relative to the center of the spectral trace and subtracted it from the region of interest. Correction of the curvature of the spectral trace was performed by shifting columns to align the center of the spectral trace along the same row. Then optimal spectral extraction as defined in \citet{horne_optimal_1986} was performed for each integration within eight pixels on both sides of the trace. The spatial profile used in the optimal extraction weighting is based on a median frame cleared of 10$\sigma$ outliers. We excluded the five spectroscopic light curves with scattering factors greater than 1.6$\times$ photon noise in the presented spectra.

\texttt{SPARTA} is a new, end-to-end pipeline that begins with the raw, uncalibrated files with \textit{JWST}. \texttt{SPARTA} has its own up-the-ramp fitting that does not depend on the JWST pipeline. The detailed reduction algorithms of \texttt{SPARTA} can be found in \citet{kempton_reflective_2023}. For F332W2 and F444W, we let the spectral center be at the 35th and 33rd pixel respectively and the extraction window to be 8 pixels. For spectroscopic light curve fitting, we excluded three data points with scattering factors greater than 1.35 or smaller than 1.0. As with \texttt{Eureka!}, the transit light curve parameters were estimated using the dynamic nested sampling algorithm \citep{Higson_2018} as implemented by the \texttt{dynesty} package\footnote{\url{https://dynesty.readthedocs.io/en/stable/}}.

We generated both ``white'' light curves that were summed over the full bandpass of each observation and spectroscopic light curves that were summed over 0.04\,$\mu m$ each (yielding 69 channels in total) for the LW data. A 9$\sigma$ outlier rejection on each light curve were performed and we fit each light curve with a transit model from the \texttt{batman} code \citep{kreidberg_batman_2015} combined with a systematics model that is linear with time (i.e., $c_0+c_1 t$). We trimmed the first 990 integrations of the first observation and 1,176 integrations of the second (both approximately the first 1 hour) due to the strong exponential-like ramp at the beginning of the observations.

We adopted the orbital period as 3.52474859\,days \citep{stassun_accurate_2017}, eccentricity as 0.0, and argument of periapsis as 90.0$\degr$. The inclination $i$ and semi-major axis in units of the host star radius $a/R_s$ were determined by fitting the white light curve of the second visit because it has less noise than the first visit (more photons were collected in these shorter wavelength data). Then $i$ and $a/R_s$ were fixed in the analysis of all the spectroscopic channels. The mid-transit time of the two observations is determined by fitting the white light curve of each observation. The best estimated parameters can be found in Table \ref{tab:lc_best_fit}. 

As is typical, the measured transit depths depend on the limb darkening assumptions. We tried using limb-darkening coefficients for a four-parameter ``non-linear'' law calculated from 3D stellar model atmospheres specific to HD\,209458 \citep{hayek_limb_2012}, but the results do not match our data well. Therefore, we elected to instead fit for the limb darkening coefficients in the light curve modeling. We adopted the quadratic limb-darkening law re-parameterised by \citet{kipping_efficient_2013}.

The SW data were reduced by enabling photometric analysis in \texttt{Eureka!} \citep[see][]{bean_high_2023}. First, the centroid of the image was determined by a 2D Gaussian fit. We then performed aperture photometry with radius of 45 pixels and a background annulus spanning from 100 to 120 pixels because this combination minimized the scatter in the light curve. We didn't correct for $1/f$ noise because it was not evident in our data. As we did for the LW data, the light curves from SW were fitted using dynamic nested sampling.

The transmission spectrum of HD\,209458b measured in the LW and SW NIRCam data by \texttt{Eureka!} is shown in Figure \ref{fig:spectrum_and_fit}. The weighted mean difference between the independent \texttt{Eureka!} and \texttt{SPARTA} reductions of the LW data is 1.67 $\sigma$ (see Figure \ref{append:comp} for the comparison). The overlapping region of the two LW filters (3.90 to 4.00 $\mu m$) agrees at 1.9 $\sigma$ for \texttt{Eureka!} reduction and 0.37$\sigma$ for \texttt{SPARTA}. However, we didn't include the SW data in the atmospheric retrieval (see next section) because of the high scatter seen in the light curves ($\sim$ 10$\times$ photon noise).

\section{Atmospheric Modeling}
\label{sec:modeling}
We retrieved the atmospheric properties of HD\,209458b by fitting the \textit{JWST} spectrum using \texttt{PLATON}\footnote{\url{https://platon.readthedocs.io/en/latest/}} \citep{zhang_forward_2019, zhang_platon_2020}, which has been used to determine the properties of several hot Jupiters \citep{jiang_evidence_2021, ahrer_lrg-beasts_2022, spyratos_precise_2023, bean_high_2023, prune_exoplanets}. Our retrieval setup assumed an isothermal atmosphere with equilibrium chemistry and a ``patchy'' grey cloud deck \citep[the latter motivated by][]{line2016}. The retrieval included the planet radius at 10$^5$\,Pa, temperature, metallicity ([M/H] = log(M/H) - log(M/H)$_{\mathrm{Sun}}$), carbon-to-oxygen (C/O) ratio, cloud-top pressure, and cloud fraction on the day-nightside terminator as free parameters. For the cloud fraction we use the prescription of \citet{pinhas_h_2o_2019}. Additionally, we included the mixing ratios of CH$_4$, NH$_3$, C$_2$H$_2$, and HCN  as free parameters (with log-uniform priors from $10^{-10}$ to $10^{-2}$) to obtain constraints on their abundances separate from the equilibrium chemistry predictions because these four species are key detections in \citet{giacobbe_five_2021}.
% because tests (see below) indicate that the \textit{JWST} spectrum has good sensitivity to these species %

Two additional cloud models are tested as follows. The first one is with spectral slope (where the absorption coefficient is characterized by ${\alpha \propto A \times \lambda^{slope}}$)\citep{zhang_forward_2019}. Besides the six parameters described above, we retrieved scattering amplitude $A$ and slope, but neither of them can be well constrained. The second is Mie scattering which has a complex refractive index of ${1.33-0.1j}$. We retrieved number density and the size of particles and the resulting C/O and [M/H] are consistent with the simpler ``patchy'' grey cloud deck model described above.

The abundance grid is computed by \texttt{FastChem}\footnote{https://github.com/exoclime/FastChem} and has five dimensions: species name (in total 34 atomic and molecular species, same as \citet{zhang_forward_2019}), temperature (100 -- 3000K with 100K interval), pressure($10^{-4}$ -- $10^{8}$ Pa with decade interval), metallicity(from [M/H] = -1 to [M/H] = 2 with 0.03 interval) and C/O (0.001 to 2.0\footnote{the interval is non-linear: [0.001, 0.005, 0.01, 0.02, 0.03, 0.04, 0.05, 0.1, 0.2, 0.4, 0.5, 0.6, 0.7, 0.8, 0.9, 0.95, 1.0, 1.05, 1.1, 1.2, 1.4, 1.6, 1.8, 2.0]}). The carbon abundance is computed by scaling the solar oxygen abundance with different C/O, and the other elements' abundances (except H and He) are scaled with metallicity.

The absorption coefficients used in the modeling are from the \texttt{DACE} opacity database\footnote{\url{https://dace.unige.ch.}} with a resolution of $R = 100,000$ over 0.5 -- 12\,$\mu$m. We included CH$_4$, CO, CO$_2$, H$_2$O, H$_2$S, NH$_3$, C$_2$H$_2$, HCN, and SO$_2$ as these are the molecules making major contribution within this wavelength range \citep{pinhas_h_2o_2019}. 
%The linelists we used are the same as described in \citet{zhang_platon_2020} Tables 4 $\&$ 10, except C$_2$H$_2$ from \citet{hitran} and CO$_2$ from UCL-4000 \citep{CO2_UCL4000}. For all of these molecules, we cut the line wings at absolute wavenumbers  (i.e. cutMode = 0) 25\,cm$^{-1}$ for P $\leq$ 10\,bar and 250\,cm$^{-1}$ for 100 and 1000\,bar, following the suggestion in \citet{line_cutoff}. 
%Cutting the line wings at 500 Lorentz widths \citep[as suggested by][]{Grimm_2015}  shows inconsistency with the 25\,cm$^{-1}$ cutoff. However, when applying the opacities with 500 Lorentz widths on WASP-39b (as described in the following paragraph), the model does not align well with the data. Therefore, we choose to keep the previous cutoffs. 
The adopted absorption coefficients that are compatible with \texttt{PLATON} used in our retrieval can be found in \S\ref{data_availability}. 

As a consistency check, we used \texttt{PLATON} to retrieve the properties of WASP-39b from the JTEC ERS NIRSpec G395H spectrum of the planet \citep{alderson_g395h_2023}. We removed the data from 3.9 $\mu$m to 4.1 $\mu$m on WASP-39b's spectrum that covers the SO$_2$ absorption feature produced by photochemistry \citep{tsai_photochemically-produced_2023}. The best fit has T = 982$^{+40}_{-39}$\,K, [M/H] = 1.39$\pm 0.16$ and C/O = 0.66$^{+0.11}_{-0.25}$, which are consistent with the results in \citet{alderson_g395h_2023} and \citet{constantinou_early_2023}. However, we found a strong degeneracy between temperature and $R_p$ (see more discussion of this in the next section). 

%Following \citet{alderson_g395h_2023}, we also enforced higher T ($>$ 1100\,K) and the derived [M/H] and C/O are consistent with theirs. 

\section{Results}
\label{sec:results}
In Figure \ref{fig:spectrum_and_fit}(a) we show the best-fit model to our \textit{JWST} transmission spectrum of HD\,209458b, with the contributions from H$_2$O, CO$_2$, CO, H$_2$S and the patchy cloud highlighted. The absorption features in the spectrum are primarily due to water (feature centered at 2.8\,$\mu m$) and carbon dioxide (feature centered at 4.3\,$\mu m$), with perhaps a minor contribution from H$_2$S. The water and carbon dioxide features are reduced in amplitude by about a factor of two due to the presence of a cloud deck. Our data favor a cloud patchiness fraction of $\sim$68\% at 3.0$\sigma$ significance\footnote{calculated as median divided by deviation}. Previous \textit{HST}/WFC3 observations taken at shorter wavelengths \citep{deming_infrared_2013-1} identified water vapour. This is the first detection of CO$_2$ in HD209458b's atmosphere, continuing the trend of detections of this molecule using JWST after WASP-39b \citep{w39_co2}, HD\,149026b \citep{bean_high_2023}, and K2-18b \citep{madhusudhan2023carbonbearing}. There is no evidence for additional absorbers.

%Our equilibrium chemistry retrievals predict log[$\chi_{\mathrm{H}_2\mathrm{O}}$] = -2.3 ($\sim$10 $\times$ solar), log[$\chi_\mathrm{CO}$] = -3.7, log[$\chi_{\mathrm{H}_2\mathrm{S}}$] = -3.9, and log[$\chi_{\mathrm{CO}_2}$] = -5.8 at the cloud top pressure P = 67 Pa.

\begin{figure}[ht!] 
\includegraphics[width = 9cm]{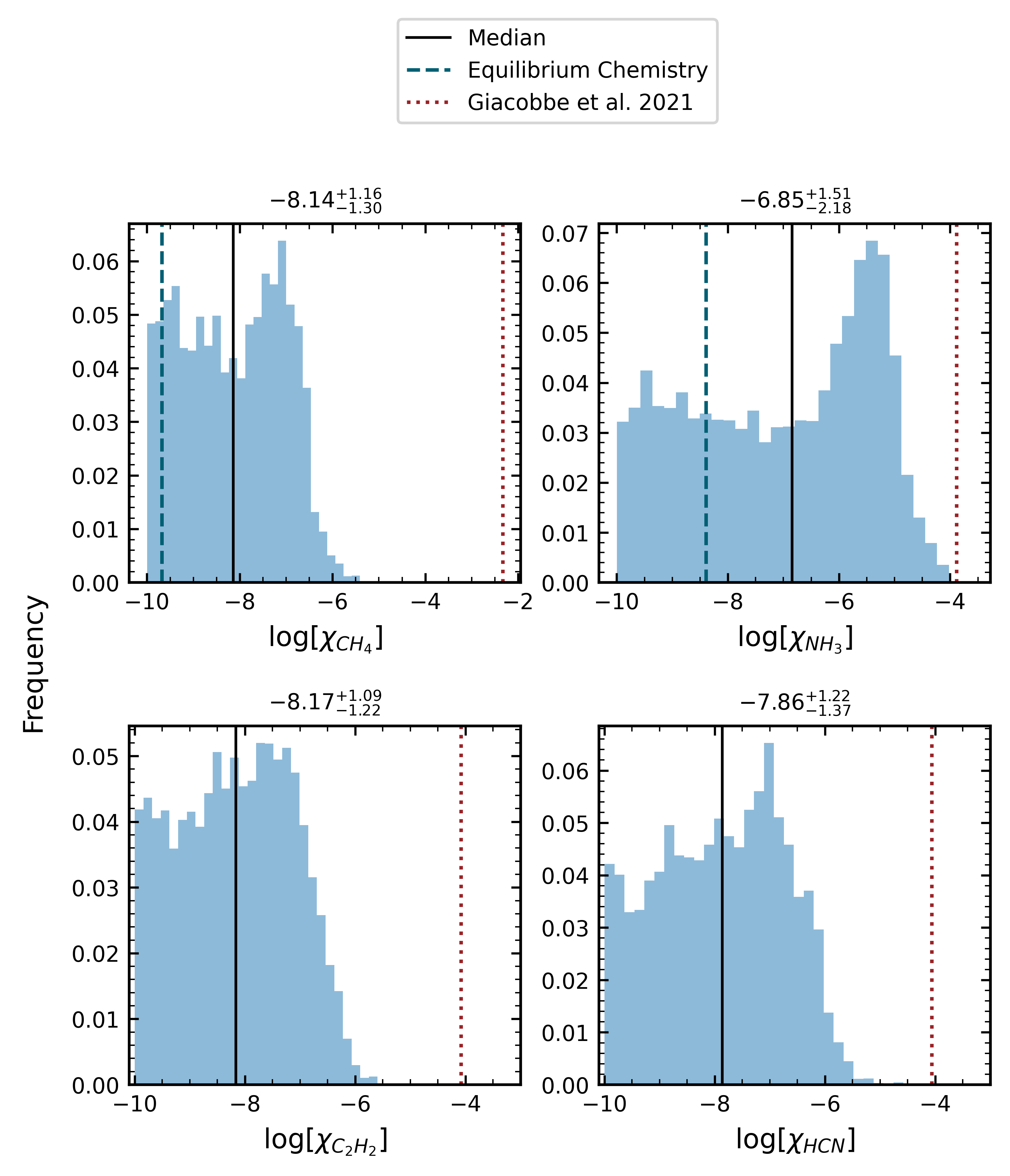} \\
\caption{Posteriors of the mixing ratios of CH$_4$, NH$_3$, C$_2$H$_2$ and HCN. The black lines show the median of the posterior. The blue dashed lines indicate the abundance predicted by equilibrium chemistry at T = 1088\,K and P = 49\,Pa (0.5\,mbar). For C$_2$H$_2$ and HCN the equilibrium values are off the plots to the left. The red dotted lines show the proposed abundances of \citet{giacobbe_five_2021}
\label{fig:free_mols}}
\end{figure}

\begin{table*}[!hpt]
  \caption{Equilibrium chemistry retrieval results.}
  \label{tab:retrieval_best_params}
  \centering
  \begin{tabular}{c c >{\bfseries}c c}
      \hline
      \multirow{2}{*}{Parameter} & \textit{HST}/WFC3 + \textit{JWST}  & \textit{JWST} only  & \textit{JWST} only \\
       & (\texttt{Eureka!}) & (\texttt{Eureka!})& (\texttt{SPARTA})\\
      
      \hline
      $R_p$ (R$_J$\footnote{Assumed Jupiter radius = $7.1492 \times 10^7$\,m}) & $1.339^{+0.007}_{-0.006}$ & 
    \boldmath{$1.353^{+0.006}_{-0.007}$} & $1.350^{+0.006}_{-0.006}$ \\
      
      T (K) & $1290^{+83}_{-81}$ & \boldmath{$1088^{+103}_{-88}$} & $1126^{+85}_{-69}$ \\
      
      [M/H] & $0.69^{+0.34}_{-0.25}$ & \boldmath{$0.54^{+0.30}_{-0.23}$} & $0.86^{+0.33}_{-0.49}$\\
      
      C$/$O & $0.23^{+0.12}_{-0.15}$ & \boldmath{$0.11^{+0.02}_{-0.06}$} & $0.06^{+0.10}_{-0.04}$ \\
      
      log10(Cloudtop Pressure) (Pa) & $1.32^{+0.45}_{-0.44}$ & \boldmath{$1.69^{+0.50}_{-0.68}$} & $1.77^{+0.38}_{-0.62}$ \\

      cloud fraction & 
      $0.82^{+0.09}_{-0.09}$&
      \boldmath{$0.68^{+0.19}_{-0.20}$} &
      $0.75^{+0.19}_{-0.28}$\\

      WFC3 offset (ppm) &
      $126^{+12}_{-11}$&/
      &/
      \\
      \hline
  \end{tabular}
\end{table*}

In Figure \ref{fig:spectrum_and_fit}(b) we present models where we scale our equilibrium abundance of CH$_4$, NH$_3$, C$_2$H$_2$, and HCN to the notional abundances from \citet[][their Extendend Data Table 4]{giacobbe_five_2021}, which gave the maximum cross-correlation signal in their data on a species-by-species basis. These molecules have absorption features in our bandpass and would have shown up (to varying degree) if their abundances were as high as those suggested by \citet{giacobbe_five_2021}. By including the volume mixing ratios of CH$_4$, NH$_3$, C$_2$H$_2$, and HCN as free parameters in our retrieval, we provide $3\sigma$ upper limits of (see Figure \ref{fig:free_mols}) log($\chi_{\mathrm{CH}_4}$) = -5.6,\, log($\chi_{\mathrm{NH}_3}$) = -4.2, log($\chi_{\mathrm{C}_2\mathrm{H}_2}$) = -5.7, and log($\chi_{\mathrm{HCN}}$) = -5.1. The posteriors for the abundances of all four molecules are consistent with the chemical equilibrium prediction from our best-fit model.

Of the four extra molecules that we explored, only the notional abundance of NH$_3$ from \citet{giacobbe_five_2021} is potentially consistent with our retrieval results i.e.,  our 3$\sigma$ upper limit is within an order of magnitude of their abundance). Our posterior for NH$_3$ is not bounded on the low end, which is consistent with the constraints for this molecule from both \citet{macdonald_hd_2017} and \citet{macdonald_signatures_2017}, which are based on \textit{HST}/WFC3 data. Therefore, the current space-based data are unable to determine if the molecule is present at equilibrium or higher abundances.

To test if the cloud prescriptions would influence the detection of CH$_4$, NH$_3$, C$_2$H$_2$, and HCN, we repeated the two cloud models described in \S\ref{sec:modeling} (paragraph 2) and recomputed the abundances of these four key molecules. We report the constrained $3\sigma$ upper limits of CH$_4$, C$_2$H$_2$, and HCN are all less than $10^{-5}$.

The abundances suggested by \citet{giacobbe_five_2021} for the other three molecules we explored (CH$_4$, HCN, and C$_2$H$_2$) are highly inconsistent with our results (i.e., our 3$\sigma$ upper limits are at least an order of magnitude lower than their abundances). However, the reported volume mixing ratios from \citet{giacobbe_five_2021} are from models that maximize the cross-correlation functions for each species individually, and thus are not retrieved values with proper uncertainties. Therefore, the spectra shown in Figure \ref{fig:spectrum_and_fit}(b) might not represent the actual inference from their data, and further analysis is needed to assess the level of agreement.

\begin{figure*}[ht!]
\plotone{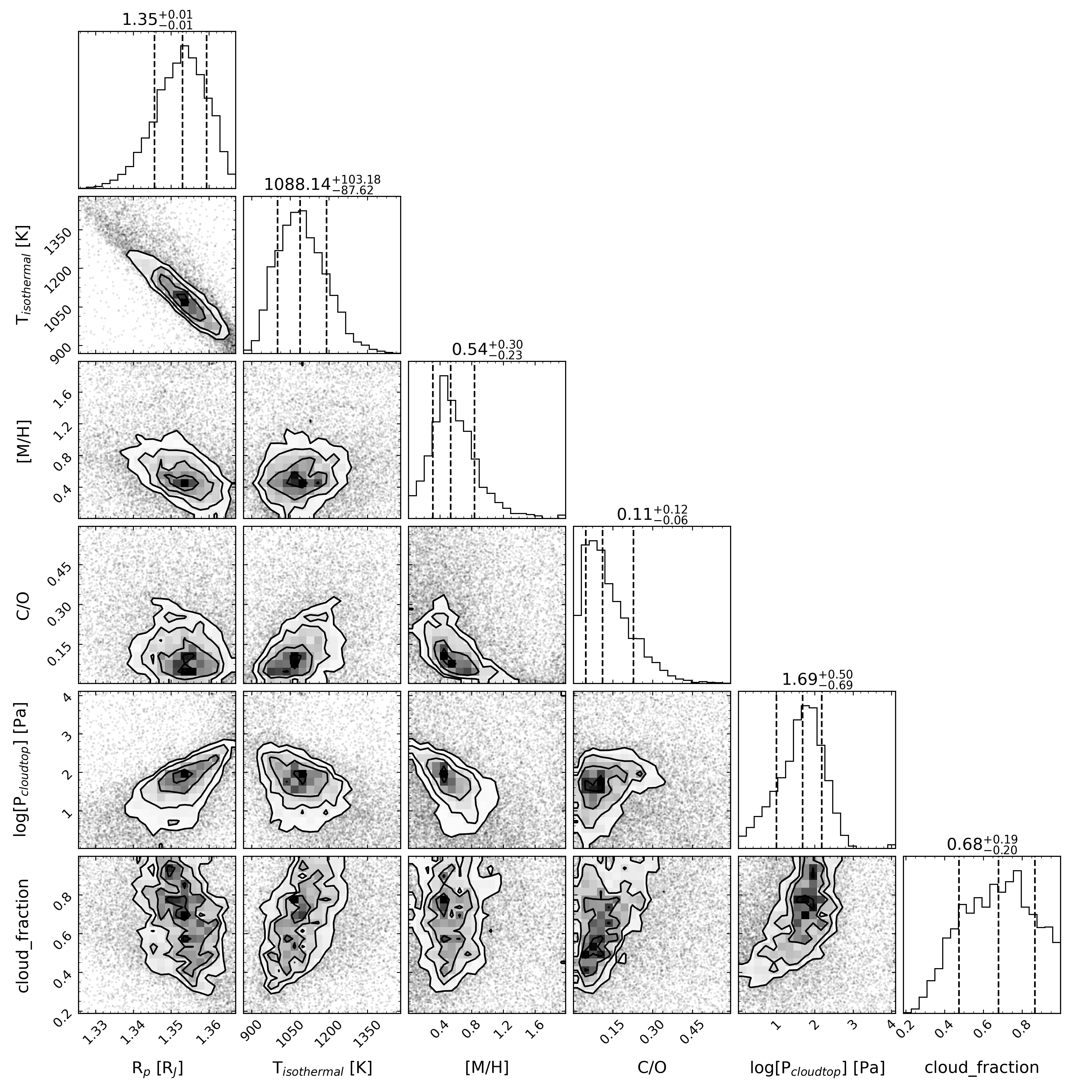} \\
\caption{Corner plot of the equilibrium chemistry retrieval. 
\label{fig:cornerplots}}
\end{figure*}

In Figure \ref{fig:cornerplots}, we show the corner plot for our chemical equilibrium retrieval on the \texttt{Eureka!} reduction. Similar to our retrieval on WASP-39b, we found a strong correlation between the temperature and the planet radius, which we believe is caused by the limitation of wavelength coverage. Specifically, the spectrum lacks the continuum fully outside molecular absorption bands and the multiple bands of the same molecule that are helpful for breaking degeneracies in transmission spectra \citep{benneke12}. On the other hand, these \textit{JWST} data precisely resolve the shape of the H$_2$O and CO$_2$ features, and our assumption of chemical equilibrium provides additional constraints on the retrieval.

To test the robustness of the wavelength coverage of our JWST data, we compared it with the best-fit parameters from joint \textit{HST}/WFC3 and \textit{JWST} fitting, and the retrieved  [M/H], C/O are consistent at greater than 0.5$\sigma$. We also conducted a test to see the impact of the minor difference in the spectrum obtained by a different data reduction pipeline. We found the retrieved parameters from \texttt{Eureka!}'s spectrum are within 1$\sigma$ of those from \texttt{SPARTA}'s spectrum. The best retrieved atmospheric properties can be found in Table \ref{tab:retrieval_best_params}.

\begin{figure}[ht!]
\centering
\plotone{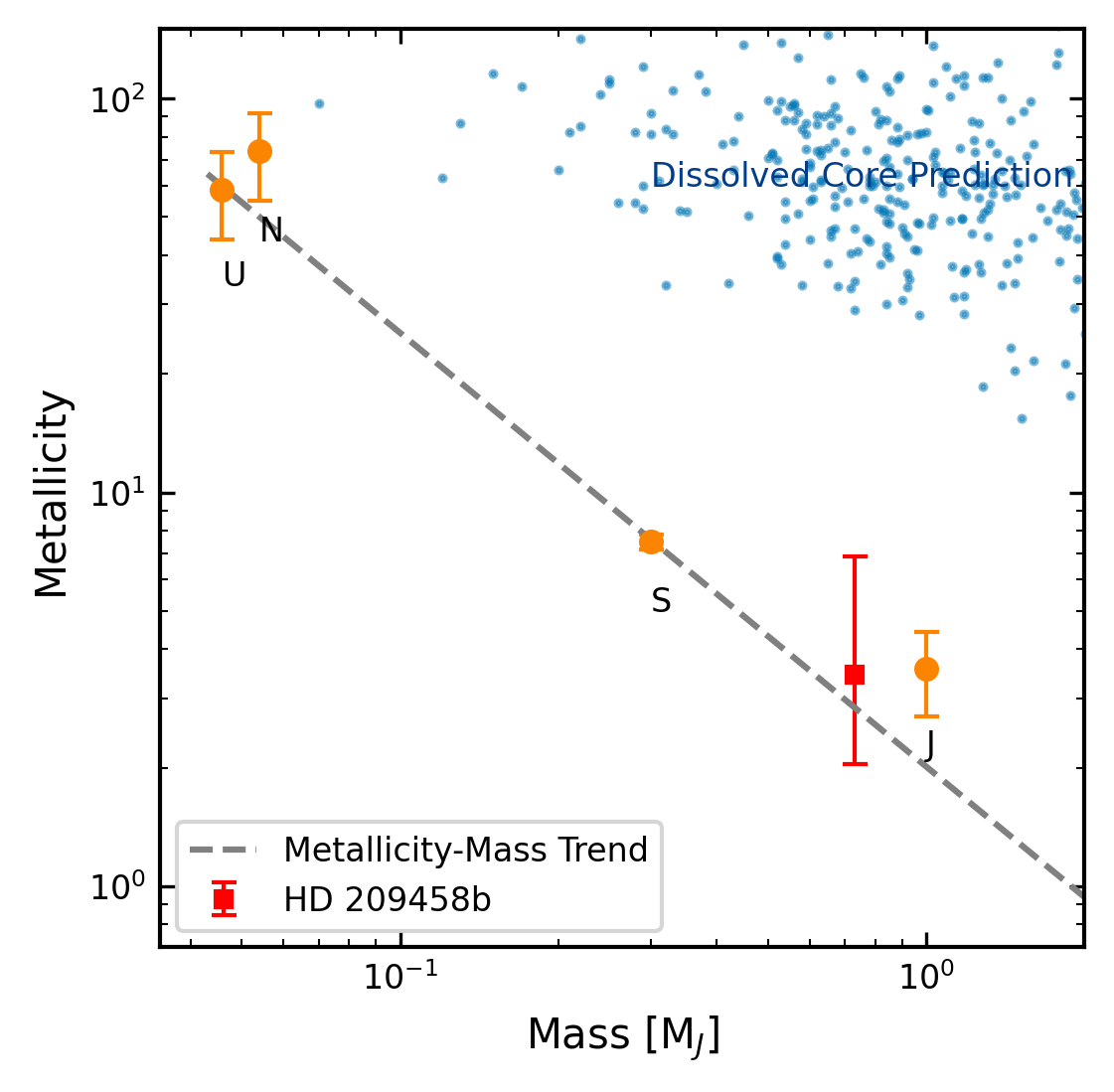} \\
\caption{Atmospheric metallicity - planet mass trend. Grey dashed line shows metallicity-mass trend linearly regressed by (M/H) relative to solar = A$\times$log(Mass) + B. The solar system giant planet data adapted from \citet{thorngren_massmetallicity_2016} with the carbon abundance adopted as a proxy for the overall metallicity. The blue circles are obtained from \citet{Thorngren_2019} showing a planet without a core with all metals uniformly mixed throughout the gas. 
\label{fig:Fe_mass}}
\end{figure}

Using only the \textit{JWST} data, \texttt{PLATON} favors a metallicity of $3^{+4}_{-1}$ $\times$ solar and a C/O = $0.11^{+0.12}_{-0.06}$ (bolded column in Table \ref{tab:retrieval_best_params}). The observed planetary atmospheric metallicity-mass trend in our solar system  has motivated a number of studies \citep{thorngren_massmetallicity_2016,welbanks_massmetallicity_2019}. As can be seen in Figure \ref{fig:Fe_mass}, our retrieved M/H is within $1\sigma$ of the trend of methane abundances in the solar system giant planets. Our findings suggest that HD\,209458b might have undergone a similar amount of planetesimal accretion as the solar system giant planets.

The interior model in \citet{Thorngren_2019} anticipates the maximum metal enrichment for an exoplanet population given its mass and radius for a ``core-less'' planet (i.e., the metals and gas are thoroughly mixed within the whole planet). These upper limits (as shown with blue circles in Figure \ref{fig:Fe_mass}) greatly exceed the observed Jupiter and Saturn metal enrichments, implying that a huge percentage of metals must be trapped within a core. The atmospheric metallicity of HD\,209458b below this limit indicates that some of the accreted metal is also in its core.

The constraint on the atmospheric C/O of HD\,209458b from our spectrum is driven by a combination of the detected H$_2$O and CO$_2$, and the lack of detection of other molecules like CH$_4$ that would be present in atmospheres with higher C/O values. In Figure \ref{fig:spectrum_and_fit}(b), we show models with C/O = 0.5, 1.0, and 1.5. We find that with a higher C/O ratio, the water feature at 2.3\,$\mu m$ is weakened and the CO$_2$ feature is enhanced until both of them are absent for C/O $>$ 1. CH$_4$ takes up the high carbon abundance above C/O ratios of unity.

\begin{figure*}[ht!]
\centering
\includegraphics[width = 18.3 cm]{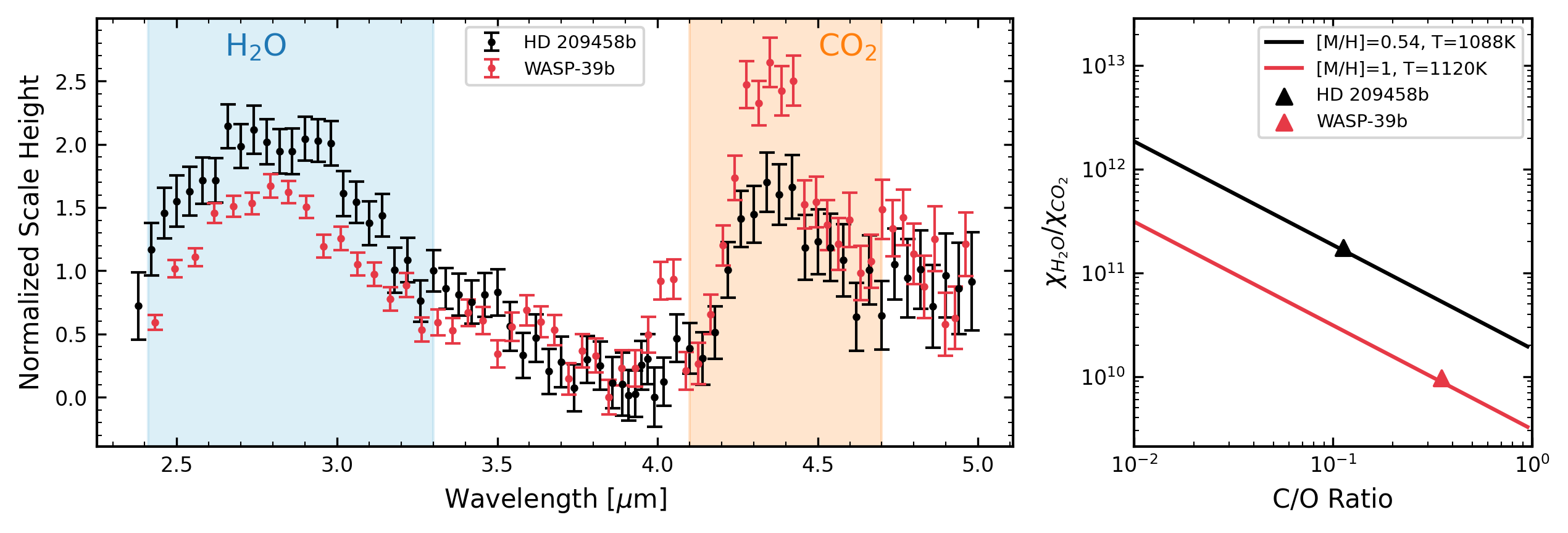} \\
\caption{Left: \textit{JWST} transmission spectra of HD\,209458b and WASP-39b \citep[spectrum adopted from][]{rustamkulov_prism_2023}) normalized by their atmospheric scale heights. Right: Calculated ratio of water and carbon dioxide abundance as a function of C/O at P = 1 mbar for the retrieved metallicity and temperature of each planet, calculated under the assumption of equilibrium chemistry. The triangle points show the abundance ratio based on their retrieved C/O. As can be seen from the spectra, HD\,209458b has a higher ratio of H$_2$O to CO$_2$ abundances, thus implying that it has a lower C/O ratio.
\label{fig:scale_heights}}
\end{figure*}

The low C/O ratio for HD\,209458b inferred in our work is due to the strong H$_2$O absorption feature, which indicates a high oxygen abundance. As a comparison, we plotted the \textit{JWST} transmission spectra in units of scale height for WASP-39b and HD\,209458b in Figure \ref{fig:scale_heights} (left). HD\,209458b exhibits a relatively stronger H$_2$O feature and a weaker CO$_2$ feature in comparison to WASP-39b. The latter exoplanet has a C/O ranging from 0.3 to 0.46, as reported in the NIRSpec G395H paper \citep{alderson_g395h_2023}. The C/O from this particular paper is chosen due to its bandpass similarity to our work. In Figure \ref{fig:scale_heights} (right), we show the ratio of H$_2$O and CO$_2$ abundances as a function of C/O ratio expected from chemical equilibrium. The $\chi_{
\mathrm{H}_2\mathrm{O}}/\chi_{\mathrm{CO}_2}$ has a dependence on metallicity because CO$_2$ itself is a strong function of metallicity. Nevertheless, given the similar metallicities of the two planets \citep[WASP-39b is $\sim$ 10 $\times$ solar,][]{w39_co2}, the $\chi_{
\mathrm{H}_2\mathrm{O}}/\chi_{\mathrm{CO}_2}$ ratio is mostly indicative of the different C/O ratios of the planets. The relative strength of H$_2$O vs.\ CO$_2$ absorption thus demonstrates that HD\,209458b's C/O ratio is significantly lower than that of WASP-39b.

% To ensure the reliability of our results, we also conducted an independent retrieval using grids of 1D radiative-convective-thermochemical equilibrium models. This approach allowed us to determine T$_p$, metallicity, and C/O and the results exhibit a high level of consistency.

In addition to the main atmospheric retrieval described above, we performed a grid fit using the ScCHIMERA Radiative Convective Equilibrium (RCE) solver first described in \citet{pis18} and recently used in JWST observations of WASP-39b \citep{rustamkulov_prism_2023}, WASP-96b \citep{Radica2023}, and WASP-80b \citep{bell2023methane}. We computed a grid of models under the assumption of 1-dimensional RCE given an irradiation and elemental composition. The grid is computed for steps in the irradiation temperature of  $T_{\rm irr}$ (1347 -- 1557\,K in steps of 15\,K), [M/H] (-0.5 -- 2.25 in steps of 0.125), and C/O (0.1 -- 0.75 in steps of 0.05). A detailed description of the ScCHIMERA grid and parameter estimation is available in \citet{Radica2023} and \citet{bell2023methane}.

Generally, we performed the parameter estimation over the ScCHIMERA grid by post-processing the 1D-RCE atmospheric structures through a transmission spectrum routine while considering the presence of inhomogeneous clouds and power-law hazes. The resulting parameter estimation derived a metallicity of $\sim 1-2\times$ solar and a sub-solar C/O ratio. The C/O ratio runs up against the lower bound limit of the grid (0.1) and has a 2$\sigma$ upper limit of 0.32. Given the consistency of the results from the two different types of retrievals we emphasize the PLATON results as the main finding in this paper.

\citet{kawashima_implementation_2021} report large differences in CH$_4$ abundance and C/O between equilibrium and disequilibrium limits when retrieving spectra covering 2.5 -- 4.0 $\mu$m. By introducing eddy diffusion transport, they found CH$_4$ quenches at P = 1 bar thus resulting in less CH$_4$ compared to the equilibrium case and C/O is raised to 0.5 in disequilibrium as compared to 0.19 in equilibrium. In order to test the impact of chemical disequilibrium on C/O, on top of the six parameters discussed in \S\ref{sec:modeling}, we added one more free parameter \texttt{P\_quenching} in \texttt{PLATON}, the pressure at which quenching happens. However, \texttt{P\_quenching} cannot be constrained according to our retrieval. Though our results disfavor the disequilibrium scenario, this may partly be because we assume the same quenching pressure for all molecules, while in reality it should be different \citep{moses_chemical_2014}. We also assume an isothermal atmosphere with fixed abundance profile while in \citet{kawashima_implementation_2021} the profile varies within the retrieval. Further modeling work on this data set using more sophisticated treatments of disequilibrium chemistry are thus warranted.

\section{Discussion}
In this paper, we present the transmission spectrum of the transiting hot Jupiter HD\,209458b with data observed with \textit{JWST}/NIRCam from 2.3 -- 5.1\,$\mu$m. The data show clear features from water, carbon dioxide, and clouds. We do not detect evidence for additional molecules that had been previously claimed for this planet. Our retrieval results suggest a mild atmospheric metallicity enhancement between that of Jupiter and Saturn in our own solar system. The data also suggest a very low C/O ratio that stands out from other emerging \textit{JWST} results for giant exoplanets.

In terms of non-detections, our upper limits on the abundances of CH$_4$, HCN, and C$_2$H$_2$ in particular make the presence of these molecules in the atmosphere of HD\,209458b as claimed by \citet[][HCN only]{hawker_evidence_2018} and \citet[][all three molecules]{giacobbe_five_2021} controversial. Nevertheless, it is important to point out that \citet{giacobbe_five_2021} did not perform a retrieval to put formal constraints on the abundances of the molecules they detected due to the challenge of such analyses on HRS data \citep[e.g.,][]{brogi19}. Therefore, it is not clear what the statistical significance is of the possible tension between the results. HRS in principle may be sensitive to trace species that elude low-resolution spectroscopy, but it remains to be seen whether the very low abundances constrained by our data would still yield a detection in HRS data.

Studies on the robustness of molecular detections by HRS of exoplanets demonstrate that some detrending methods may induce false positive or inflated detections \citep{cheverall_robustness_2023}. In their case study on HD\,209458b, HCN, NH$_3$ and CH$_4$ were not observed by \citet{cheverall_robustness_2023}. However, a signal for CH$_4$ can be detected if the telluric contamination is not correctly removed. While methane (CH$_4$) is ubiquitous in the atmosphere of solar system giant planets, it absence has long been one of the core puzzles of the study of exoplanetary atmospheres \citep{a_new_gibson_2011, Benneke_2019, Evidence_diseq_2021, w39_co2}. However, recent \textit{JWST} observations on transiting exoplanets WASP-80\,b and K2-18\,b show evidence of methane \citep{bell2023methane, madhusudhan2023carbonbearing}. Understanding this molecule's absence or existence is essential for understanding how planets form and evolve, how atmospheric processes work, and the habitability of planets. Future JWST observations, like cycle 2 program 3557, might aid in solving this mystery.

%Previous studies using emission spectroscopy\citep{line_no_2016} and HRS \citep{giacobbe_five_2021} showed evidence for CO$_2$ but were not robust enough to claim a detection. Our clear detection of this molecule demonstrates that HD\,209458b's atmosphere is enriched in heavy elements relative to solar. 
Our study disproves the previous claims of low water abundance for this planet \citep{madhusudhan_h_2014,macdonald_hd_2017, brogi_framework_2017, welbanks_mass-metallicity_2019, pinhas_h_2o_2019}. Instead of being depleted in water, our best-fit model gives enhanced metallicity and low C/O, indicating the atmosphere of HD\,209458b is rich in oxygen. We found a metallicity that is consistent with, but more precise than \citet{line_no_2016} in emission and \citet{Welbanks2021} in transmission, which extends the agreement that is seen between the solar system trend and exoplanet atmosphere abundances.

\citet{line_no_2016} and \citet{brogi_framework_2017} have constrained the C/O of HD\,209458b to $<1$, yet our very low C/O ratio provides valuable insights into the planetary formation and evolution. \citet{espinoza_metal_2017} suggests a low C/O in gas giants compared to parent stars is caused by metal enrichment but not dependent on the formation location. Additionally, planetesimal pollution \citep{_berg_2011} caused by formation and migration sufficiently inward of the snowlines of carbon-bearing species may also result in low ($<0.5$) C/O and elevated metal enrichment. 
Through a comparative analysis of the spectra of this study and WASP-39b, we found the two have similar molecular features (namely water and CO$_2$), while the variation in the ratio of water to carbon dioxide abundance leads to a distinct difference in the C/O. This is primarily because the value of $\chi_{\mathrm{H}_2\mathrm{O}}/\chi_{\mathrm{CO}_2}$ serves as a robust indicator of C/O, assuming comparable metallicity. It is important to note that this relationship is not influenced by specific retrieval models, thus revealing the intrinsic characteristics of such planets. 

\citet{tsai_photochemically-produced_2023} showed evidence of photochemically-produced SO$_2$ in the atmosphere of WASP-39b. Since we do not detect SO$_2$ at 4.05 $\mu$m, we have not made an effort to study the potential impact of chemical disequilibrium on the H$_2$S and SO$_2$ abundance. Detailed modeling to assesss our JWST transmission spectrum in the context of the disequilibrium process would be interesting.

\section{Data Availability}
\label{data_availability}
The data presented in this paper were obtained from the Mikulski Archive for Space Telescopes (MAST) at the Space Telescope Science Institute. The specific observations analyzed can be accessed via \dataset[DOI: 10.17909/f5j3-jq48]{https://doi.org/10.17909/f5j3-jq48}. The data that were used to create all of the figures will be freely available on Zenodo \citep{xue_2024_10557924}. All additional data is available upon request.
\newpage

%% IMPORTANT! The old "\acknowledgment" command has be depreciated. It was
%% not robust enough to handle our new dual anonymous review requirements and
%% thus been replaced with the acknowledgment environment. If you try to 
%% compile with \acknowledgment you will get an error print to the screen
%% and in the compiled pdf.
%% 
%% Also note that the akcnowlodgment environment does not support long amounts of text. If you have a lot of people and institutions to acknowledge, do not use this command. Instead, create a new \section{Acknowledgments}.
\section{Acknowledgments}
\begin{acknowledgments}
We thank Matteo Brogi for helpful discussions about the high-resolution spectroscopy results for HD\,209458b. This work is based on observations made with the NASA/ESA/CSA James Webb Space Telescope. The data were obtained from the Mikulski Archive for Space Telescopes at the Space Telescope Science Institute, which is operated by the Association of Universities for Research in Astronomy, Inc., under NASA contract NAS 5-03127 for JWST. These observations are associated with program GTO 1274.

This publication makes use of The Data \& Analysis Center for Exoplanets (DACE), which is a facility based at the University of Geneva (CH) dedicated to extrasolar planets data visualisation, exchange and analysis. DACE is a platform of the Swiss National Centre of Competence in Research (NCCR) PlanetS, federating the Swiss expertise in Exoplanet research. The DACE platform is available at \url{https://dace.unige.ch.}
\end{acknowledgments}

%% To help institutions obtain information on the effectiveness of their 
%% telescopes the AAS Journals has created a group of keywords for telescope 
%% facilities.
%
%% Following the acknowledgments section, use the following syntax and the
%% \facility{} or \facilities{} macros to list the keywords of facilities used 
%% in the research for the paper.  Each keyword is check against the master 
%% list during copy editing.  Individual instruments can be provided in 
%% parentheses, after the keyword, but they are not verified.

\facilities{JWST(NIRCam)}

%% Similar to \facility{}, there is the optional \software command to allow 
%% authors a place to specify which programs were used during the creation of 
%% the manuscript. Authors should list each code and include either a
%% citation or url to the code inside ()s when available.

\software{Eureka! \citep{bell_eureka_2022},  
          PLATON \citep{zhang_platon_2020}, 
          SPARTA \citep{kempton_reflective_2023},
          Astropy \citep{astropy:2013, astropy:2018, astropy:2022},
          dynesty \citep{Speagle_dynesty_2020},
          batman \citep{kreidberg_batman_2015},
          hitran \citep{hitran},
          HELIOSK \citep{heliosk},
          Fastchem \citep{Stock_fastchem_2018}
          }

%% Appendix material should be preceded with a single \appendix command.
%% There should be a \section command for each appendix. Mark appendix
%% subsections with the same markup you use in the main body of the paper.

%% Each Appendix (indicated with \section) will be lettered A, B, C, etc.
%% The equation counter will reset when it encounters the \appendix
%% command and will number appendix equations (A1), (A2), etc. The
%% Figure and Table counter will not reset.

\appendix
\begin{figure*}[ht!]
\includegraphics[width = 16 cm]{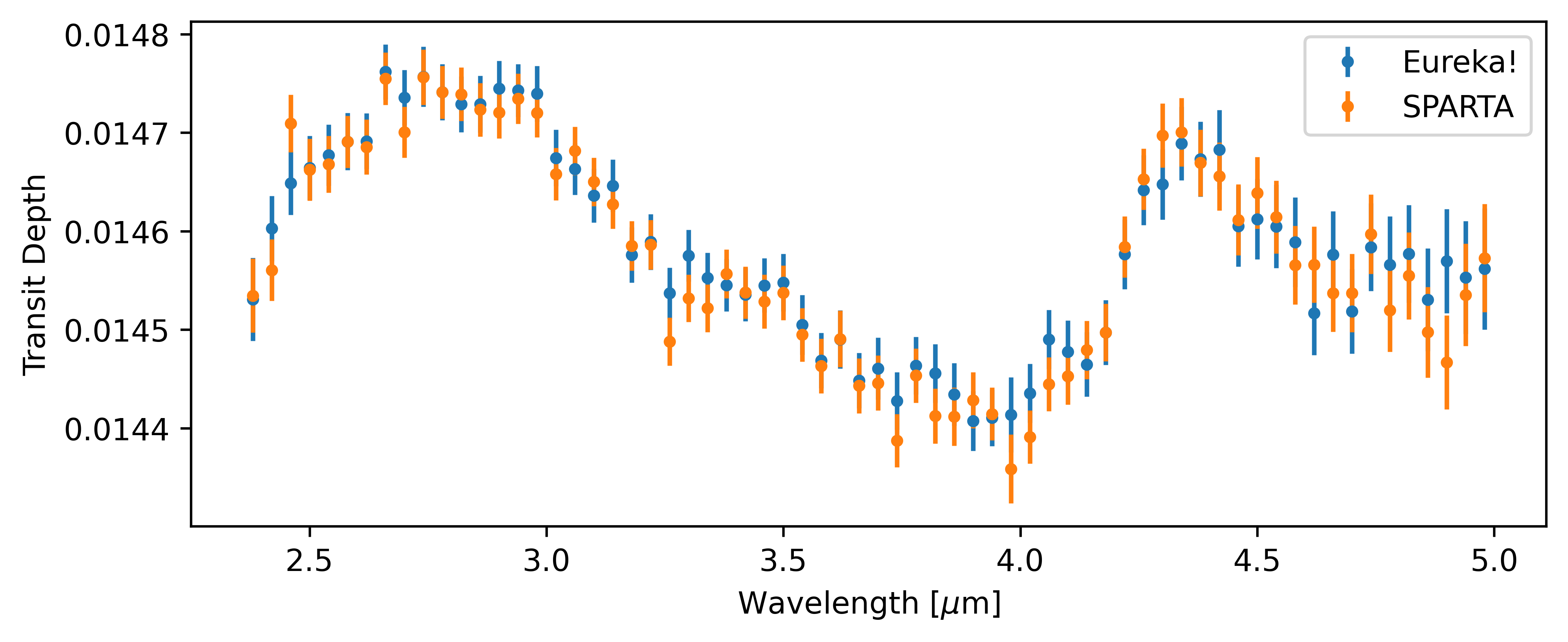} \\
\caption{Spectra reduced by \texttt{Eureka!} and \texttt{SPARTA}. The data behind the figure can be found in \S\ref{data_availability}.
\label{append:comp}}
\end{figure*}

\bibliography{hd209pub}{}
\bibliographystyle{aasjournal}

%% This command is needed to show the entire author+affiliation list when
%% the collaboration and author truncation commands are used.  It has to
%% go at the end of the manuscript.
%\allauthors

%% Include this line if you are using the \added, \replaced, \deleted
%% commands to see a summary list of all changes at the end of the article.
%\listofchanges

\end{document}